# Finite-temperature magnetism in bcc Fe under compression

Xianwei Sha* and R. E. Cohen

Carnegie Institution of Washington, 5251 Broad Branch Road, NW, Washington, D. C. 20015, U. S. A.

We investigate the contributions of finite-temperature magnetic fluctuations to the thermodynamic properties of bcc Fe as a function of pressure. First, we apply a tight-binding total-energy model parameterized to first-principles linearized augmented planewave computations to examine various ferromagnetic, anti-ferromagnetic, and noncollinear spin spiral states at zero temperature. The tight-binding data are fit to a generalized Heisenberg Hamiltonian to describe the magnetic energy functional based on local moments. We then use Monte Carlo simulations to compute the magnetic susceptibility, the Curie temperature, heat capacity, and magnetic free energy. Including the finite-temperature magnetism improves the agreement with experiment for the calculated thermal expansion coefficients.

* Now at the National Institute of Standards and Technology, 100 Bureau Drive, Gaithersburg, MD 20899.

The accurate description of materials properties at finite temperatures using first-principles based techniques remains a challenging research topic in solid state theory. Several approaches have been developed, including the quasi-harmonic lattice dynamics,[1] *ab initio* molecular dynamics,[2] particle-in-a-cell method,[3] and path integral Monte Carlo,[4] with each of them having its own advantages and disadvantages.[5] We previously used density functional perturbation theory and quasi-harmonic lattice dynamics to calculate the thermal equation of state and thermoelasticity of bcc iron. Those computations neglected magnetic fluctuations, and we found that some of the calculated thermal properties such as the thermal expansivity showed relatively large discrepancies to experiment.[6,7] In contrast, the calculated thermal expansion coefficients of nonmagnetic bcc vanadium using the same theoretical techniques agree well with experiment.[8] Kormann et al. reported that the differences between their *ab initio* Helmholtz free energy of bcc Fe and CALPHAD data obtained with the Thermocalc program and the SGTE unary database increase with temperature when the magnetic contribution is neglected.[9] Thus it is important to include the proper theoretical treatment of finite-temperature magnetism in order to accurately describe the thermodynamics of magnetic materials.

Significant progress has been achieved to model the finite-temperature magnetism.[10] Rosengaard and Johansson proposed a simple model to provide a unified description of the energetics of moment formation as well as the energetics of moment ordering, and they calculated the finite-temperature magnetic properties of bcc Fe and fcc Co and Ni using Monte Carlo simulations.[11] The same theoretical techniques have been successfully applied to tetragonal iron,[12] binary alloys,[13] and low-dimensional magnetic systems.[14] Several other theoretical approaches have also been developed, such as dynamical mean-field theory,[15] Wannier-function approach,[16] and a classical spin-fluctuation model.[17]

In this letter, we present our results on finite-temperature magnetism of bcc Fe under pressure, using similar theoretical techniques as Rosengaard and Johansson[11] with several differences. First, we examine the magnetic properties at both ambient and high pressures; second, we obtain the parameters in the generalized Heisenberg Hamiltonian via a single global fitting, instead of treating the on-site and interatomic exchange terms separately; third, we use a tight-binding total-energy model parameterized to first-principles linearized augmented plane-wave (LAPW) generalized gradient approximation (GGA) computations to examine various ferromagnetic, anti-ferromagnetic, and noncollinear spin spiral states at zero temperature. GGA is also widely known to show significant improvements in describing many ground-state properties of iron over the local density approximation (LDA) used by Rosengaard and Johansson.[18]

The Helmholtz free energy F(V, T) of many metals and alloys has several major contributions:[19]

$$F(V,T)=E_{static}(V)+F_{el}(V,T)+F_{vib}(V,T)+F_{mag}(V, T) \qquad (1)$$

where V is volume and T is temperature. We previously calculated the zero-temperature energy of a static lattice $E_{static}$, the electronic thermal free energy $F_{el}$, and the lattice vibrational energy $F_{vib}$ for ferromagnetic bcc Fe at several selected volumes and temperatures.[7] Here we use a multi-scale modeling approach to obtain the finite-temperature magnetic fluctuation energy $F_{mag}$. We first calculate the magnetic energies for a large number of ferromagnetic, antiferromagnetic and noncollinear planar spin spiral states of bcc Fe at zero temperature using tight-binding total energy fixed spin moment computations. We fit the calculated magnetic data to a generalized Heisenberg Hamiltonian to describe the magnetic energy functional based on local spin moment, and then perform extensive Monte Carlo simulations to obtain the magnetic properties as a function of temperature. The integration of the magnetic heat capacity over temperature gives $F_{mag}$.

The total energy in the tight-binding model for a magnetic material is given by

$$E = E_{bs} + \frac{1}{4}\sum_{jLL'} |m_{JL}| I_{JLjL'} |m_{jL'}| + \sum_j b_j \bullet m_j \quad (1)$$

where $I_{jLjL'}$ is the Stoner parameter that represents the exchange interaction on state L from state L' on atom j, $b_j$ is an applied magnetic field on atom j, $|m_{jL}|$ is the magnitude of the magnetic moment of state L on atom j.[20] The model is parameterized to a large number of LAPW GGA computations for different crystal structures and magnetic states of Fe and has been successfully applied to model the compression, electronic structure, phase relations and elasticity.[20, 21] We examine bcc Fe at volumes of 60, 70, 75, 79.6 and 85 bohr$^3$/atoms; 79.6 bohr$^3$/atoms is the ambient equilibrium volume. At each volume, we performed 80 independent calculations with moment varying from 0 to 4.0 for bcc Fe at constrained ferromagnetic and anti-ferromagnetic states, and additional 30 non-collinear planar spin spiral states along high symmetry directions. Tight-binding calculations of large super-cell systems including 128 and 256 atoms have been included to properly represent the longitudinal spin-fluctuation energy. In Fig. 1 we show the calculated magnetic energies and moment for planar spin spiral states along the [001] direction. The tight-binding total-energy data depend on the Stoner parameter I, and we check the results for I varying from 0.90 to 1.03. The magnetic energy and moment of the spin spiral states strongly depends on the Stoner parameter, especially at large wave vectors. The calculated data at I=0.95 give the best agreements with previous LDA results.[11] We further compare the tight-binding data of magnetic moment and energy of bcc Fe at ferromagnetic ground states with varying Stoner parameters to first-principles linear muffin-tin orbital calculations using different exchange correlation functionals, and find that the tight-binding values agree best with LDA results at I=0.95, and with the Perdew-Burke-Ernzerhof (PBE) GGA results at I=0.97.

We perform a global least-squares fitting of the calculated tight-binding magnetic energy as a function of the local spin moments to obtain the on-site and interatomic exchange parameters in the generalized Heisenberg Hamiltonian.[11] The accuracy of the fitting is shown by the reproduction of the tight-binding magnetic energies at various states from the parameterized Hamiltonian. At all volumes, the nearest-neighbor Heisenberg term $J_1$ is the dominate factor in the interatomic interaction. We carefully checked the total magnetic energy as well as the on-site and interatomic exchange contributions for bcc Fe at several constrained ferromagnetic states using the Hamiltonian, and our data at the ambient equilibrium volume are in reasonable agreements with Rosengaard and Johansson's previous results.[11]

At each volume, we perform extensive Monte Carlo (MC) simulations to obtain the magnetic properties as a function of temperature. We carefully check the convergence of the MC results relative to the system size and other factors. We obtain the temperature dependences of the magnetic moment and the inverse ferromagnetic suspsceptibility[15] from the MC simulations, as shown in Fig. 2, which further gives the Curie temperature $T_c$. As the temperature approaches $T_c$, the ordered magnetic moment drops rapidly to zero. The inverse of the uniform susceptibility shows a linear increase with temperature above $T_c$, consistent with previous dynamic mean-field theory.[15] As shown in Fig. 3, the calculated $T_c$ strongly depends on the Stoner parameter I. Our calculated $T_c$ gives the best agreement with experiment at I=0.95, in agreement with Rosengaard and Johansson whose ambient-pressure MC simulation gives the Curie temperature in excellent agreement with the experiment for bcc Fe.[11] As previously discussed, our tight-binding magnetic energies agree best with their DFT-LDA results at I=0.95. At I=0.97 where the tight-binding energies match best with DFT-GGA results, the calculated $T_c$ is ~100K higher. A much higher $T_c$ obtained using GGA functional over LDA functional has been previously reported by Kormann et al.[22] In contrast, dynamical mean-field theory gives a much higher $T_c$ of 1900K because of the failure in capturing the reduction of $T_c$ due to long wavelength spin waves and the mean field approximation.[15] Our calculated Curie temperature shows a slightly linear decrease with the decrease of atomic volume (increase of pressure according to the equation of state). Legar et al. reported experiments that showed the Curie temperature of bcc iron to be essentially pressure independent.[23] However, their measurements are limited to very low pressure up to 1.75 GPa. Moran et al.[24] and Kormann et al.[22] examined the volume dependences of the Curie temperature using LDA and GGA approximations for exchange-correlation functional, respectively. Instead of using MC simulations, they use conventional analytical solutions for the Heisenberg model with mean-field (MF) or random-phase approximation (RPA). Unfortunately, these methods are not very accurate and Rosengaard and Johansson reported that the calculated $T_c$ using MF approximation is 400 K higher than the MC results when using the same Heisenberg Hamiltonian for bcc

Fe.[11] The deviation between the calculated $T_c$ using MF and RPA approximations is as large as ~500K, and two LDA calculations using the same theoretical methods give significantly different pressure dependences of $T_c$.[22, 24]

We calculate the magnetic free energy contributions via the integration of the magnetic heat capacity over temperature obtained from MC runs. We use I=0.97 since it gives the best agreement for magnetic properties to PBE GGA calculations. The finite-temperature magnetic fluctuation energies are added to our previous PBE GGA results for the electronic and lattice vibrational contributions to obtain the Helmholtz free energies. The free energies are further fit to the Vinet equation of state to obtain various thermodynamic properties.[7] As shown in Fig. 4, the calculated thermal expansion coefficients are in much better agreement with experiment[25, 26] when the finite-temperature magnetism is taken into account. Including the magnetic fluctuation energy also gives a kink in thermal expansion around the Curie temperature, consistent with the experiment.

To conclude, we use multi-scale theoretical techniques to investigate the magnetic properties in bcc Fe as a function of pressure and temperature. The calculated values such as the Curie temperature $T_c$ show strong dependences on the Stoner parameter used in the tight-binding total-energy model. $T_c$ shows a slight decrease with the increase of pressure. Including the magnetic fluctuation contributions significantly improves the agreements of finite-temperature thermal expansion coefficients with the experiment.

**Acknowledgements**

This work was supported by DOE ASCI/ASAP subcontract B341492 to Caltech DOE w-7405-ENG-48 and by NSF grant EAR-0738061, the Carnegie Institution of Washington, and EFree, an Energy Frontier Research Center funded by the U.S. Department of Energy, Office of Science, Office of Basic Energy Sciences under Award Number DE-SC0001057.. Computations were performed at the Geophysical Laboratory and on ALC at Lawrence Livermore National Lab.

# REFERENCES

1. D. G. Isaak, R. E. Cohen and M. J. Mehl, J. Geophys. Res.-Solid Earth **95** (B5), 7055-7067 (1990).
2. A. R. Oganov, J. P. Brodholt and G. D. Price, Nature **411** (6840), 934-937 (2001).
3. E. Wasserman, L. Stixrude and R. E. Cohen, Phys. Rev. B **53** (13), 8296-8309 (1996).
4. B. Militzer, D. M. Ceperley, J. D. Kress, J. D. Johnson, L. A. Collins and S. Mazevet, Phys. Rev. Lett. **87** (27), 275502 (2001).
5. X. W. Sha and R. E. Cohen, Phys. Rev. B **74** (6), 064103 (2006).
6. X. W. Sha and R. E. Cohen, Phys. Rev. B **74** (21), 214111 (2006).
7. X. W. Sha and R. E. Cohen, Phys. Rev. B **73** (10), 104303 (2006).
8. X. Sha and R. E. Cohen, presented at the Materials Research Society Symposium Proceedings, Boston, MA, 2007 (unpublished).
9. F. Kormann, A. Dick, B. Grabowski, B. Hallstedt, T. Hickel and J. Neugebauer, Phys. Rev. B **78** (3), 033102 (2008).
10. I. Turek, J. Kudrnovsky, V. Drchal and P. Bruno, Philos. Mag. **86** (12), 1713-1752 (2006).
11. N. M. Rosengaard and B. Johansson, Phys. Rev. B **55** (22), 14975-14986 (1997).
12. J. T. Wang, D. S. Wang and Y. Kawazoe, Appl. Phys. Lett. **88** (13), 132513 (2006).
13. P. James, O. Eriksson, B. Johansson and I. A. Abrikosov, Phys. Rev. B **59** (1), 419-430 (1999).
14. S. Polesya, O. Sipr, S. Bornemann, J. Minar and H. Ebert, Europhys. Lett. **74** (6), 1074-1080 (2006).
15. A. I. Lichtenstein, M. I. Katsnelson and G. Kotliar, Phys. Rev. Lett. **87** (6), 067205 (2001).
16. E. Sasioglu, A. Schindlmayr, C. Friedrich, F. Freimuth and S. Blugel, Phys. Rev. B **81** (5), 054434 (2010).
17. A. L. Wysocki, J. K. Glasbrenner and K. D. Belashchenko, Phys. Rev. B **78** (18), 184419 (2008).
18. L. Stixrude, R. E. Cohen and D. J. Singh, Phys. Rev. B **50** (9), 6442-6445 (1994).
19. O. L. Anderson, *Equations of state of solids for geophysics and ceramic science*. (Oxford University Press, New York, 1995).
20. R. E. Cohen and S. Mukherjee, Phys. Earth Planet. Inter. **143-144**, 445-453 (2004).
21. R. E. Cohen, L. Stixrude and E. Wasserman, Phys. Rev. B **56** (14), 8575-8589 (1997).
22. F. Kormann, A. Dick, T. Hickel and J. Neugebauer, Phys. Rev. B **79** (18), 184406 (2009).
23. J. M. Leger, Lorierss.C and B. Vodar, Phys. Rev. B **6** (11), 4250-4261 (1972).
24. S. Moran, C. Ederer and M. Fahnle, Phys. Rev. B **67** (01), 012407 (2003).
25. D. E. Gray, *American institute of physics handbook*, 3. rev. ed. (McGraw-Hill, New York, 1972).
26. I. S. Grigoriev and E. Z. Meilikhov, *Handbook of physical quantities*. (CRC Press, Boca Raton, Fla., 1997).

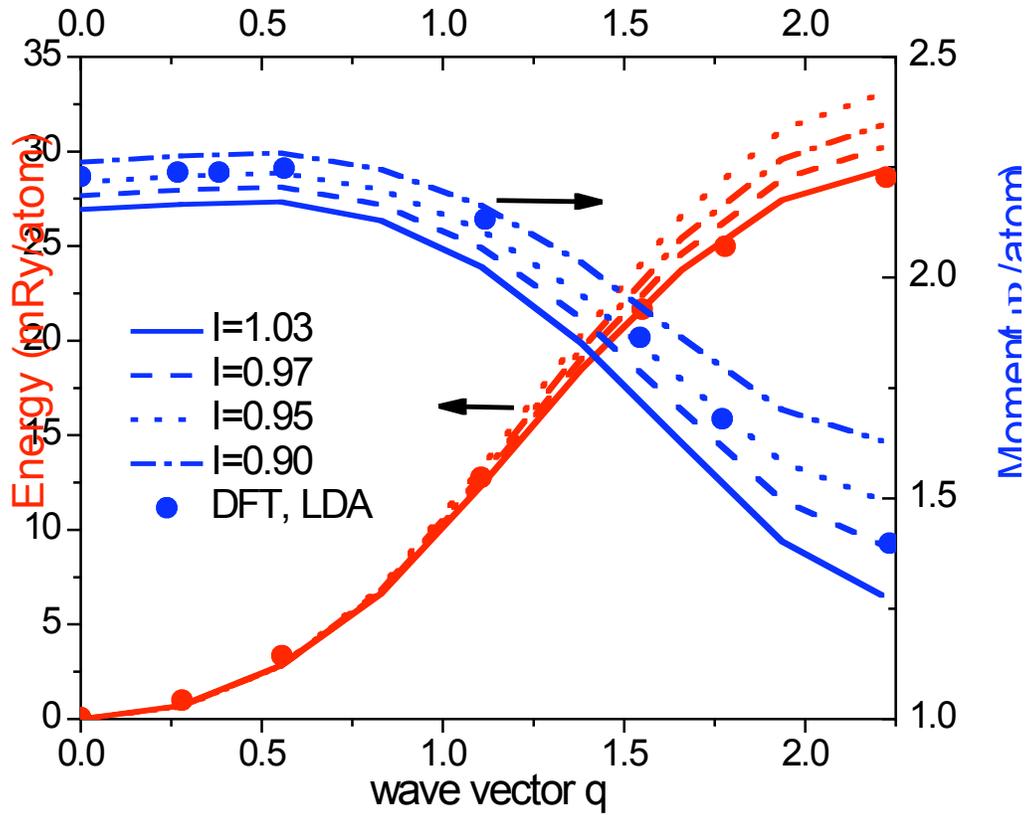

FIG 1 The calculated equilibrium local magnetic moment and total energies relative to the ferromagnetic ground state for bcc Fe in the planar spin spiralrs along the [001] direction. DFT results using LDA approxiatmions (Ref.[11]) are also shown for comparisions.

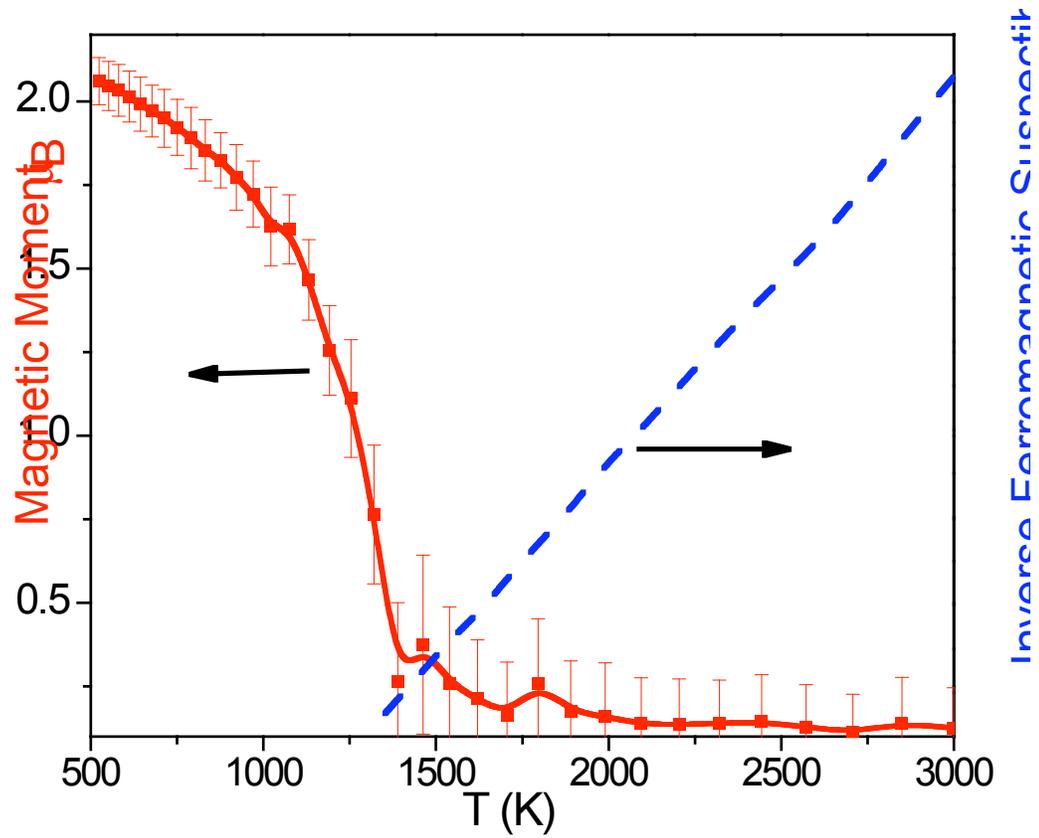

FIG. 2 The temperature dependences of the average magnetic moment and the inverse ferromagnetic susceptibility for bcc Fe obtained from Monte Carlo simulations based on the generalized Heisenberg Hamilton parameterized to the tight-binding total energy data.

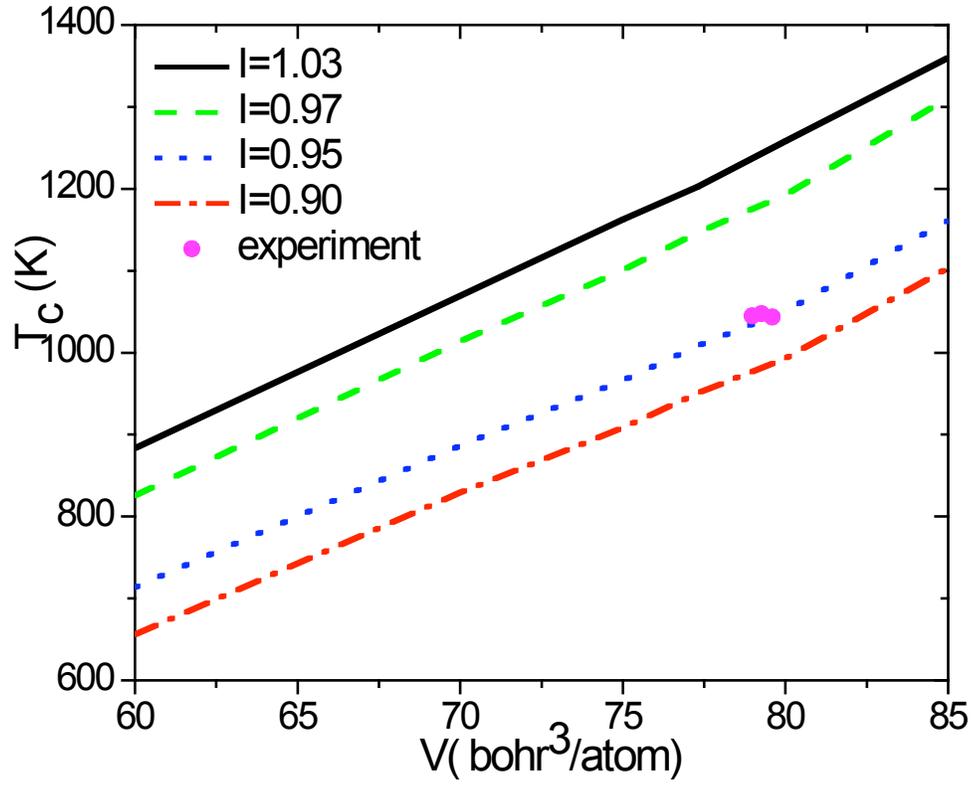

FIG 3 The calculated volume dependences of the Curie temperatures at different Stoner parameter I, in comparison to the experiment data (Ref. [23]).

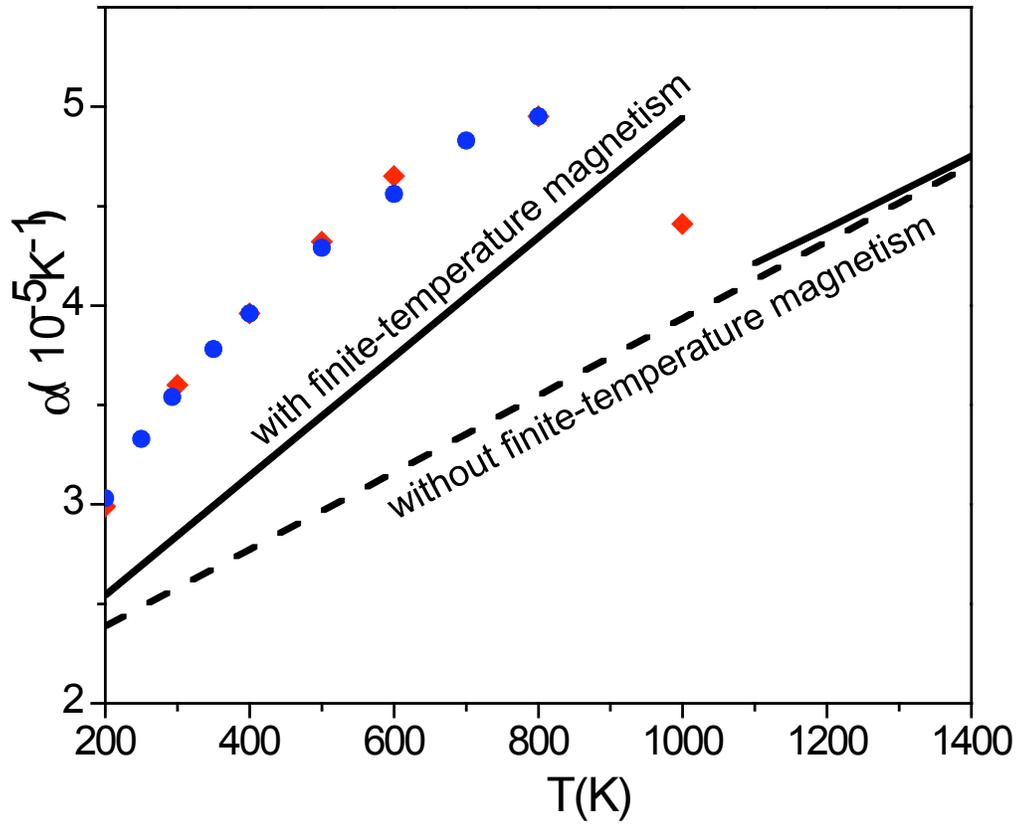

Fig. 4 The calculated ambient-pressure thermal expansion coefficients (lines), with and without finite-temperature magnetism considered. Including the magnetic fluctuation contributions to the free energy gives better agreement with experiment (filled symbols, Ref.[25, 26] ), and correctly predicts the kink around the Curie temperature in the thermal expansivity.